%% file: SecureCA.tex
\documentclass[conference]{IEEEtran}
\makeatletter
\def\ps@headings{%
\def\@oddhead{\mbox{}\scriptsize\rightmark \hfil \thepage}%
\def\@evenhead{\scriptsize\thepage \hfil \leftmark\mbox{}}%
\def\@oddfoot{}%
\def\@evenfoot{}}
\makeatother
\usepackage{url}
\usepackage[dvips]{graphicx}
\usepackage{graphics}
\usepackage{cite}
\usepackage{amssymb}
\usepackage{amsmath}
\usepackage{subfigure}
\usepackage{amsfonts}         % n?cesaire pour les caract?res blackboard (mathbb)
\usepackage{color}            % importation depuis Xfig
\usepackage{epic}             % compl?mente la suite
\usepackage{eepic}            % remplissage Xfig
\usepackage{rotating}         % texte dans les images
\usepackage{type1cm}          % polices de taille arbitraire
\usepackage{epsfig}
\usepackage{amsthm}
\usepackage{amsmath}
\usepackage{graphicx}
\usepackage{graphics}
\usepackage{epsfig}
\usepackage{latexsym}
\usepackage{amsfonts}
\usepackage{paralist}
\usepackage{xspace}
\usepackage{mathrsfs}
\usepackage{psfrag}
\usepackage{color}
\usepackage[ruled]{algorithm}
\usepackage[noend]{algorithmic}

\newtheorem{definition}{Definition}
\newtheorem{theorem}{Theorem}
\newtheorem{lemma}[theorem]{Lemma}

\def\signer{{\mathcal{T}}}
\def\auctioneer{{\mathcal{A}}}
\def\bidder{{\mathcal{B}}}

\def\zgroup{{\mathbb{Z}}}

\newcommand{\CUT}[1]{{}}

\begin{document}

\title{Enabling Privacy-preserving Auctions in Big Data}

\author{Taeho Jung$^1$, Xiang-Yang Li$^{12}$
%, Lan Zhang$^{\mathcal{z}}$, He Huang$^\mathcal{x}$
\\$^1$Department of Computer Science, Illinois Institute of Technology, Chicago, IL
\\$^{2}$Department of Computer Science and Technology, TNLIST, Tsinghua University, Beijing\\
%$^\mathcal{x}$School of Computer Science, Soochow University, Suzhou
}

\maketitle

%\footnotetext[1]{The research of authors is partially supported by NSF CNS-0832120, NSF CNS-1035894, NSF ECCS-1247944, National Natural Science Foundation of China under Grant No. 61170216, No. 61228202, China 973 Program under Grant No.2011CB302705.}

\input{abstract_v1.tex}

\input{intro_v1.tex}

\input{background_v1.tex}

\input{model_v1.tex}

\input{preliminary_v2.tex}

\input{construction_v2.tex}

\input{evaluation_v1.tex}

\input{performance_v1.tex}

\input{conclusion_v1.tex}

\bibliographystyle{ieeetr}
\bibliography{SecureCA}

\end{document}

%% file: abstract_v1.tex
\begin{abstract}
We study how to enable auctions in the big data context to solve many upcoming data-based decision problems in the near future. We consider the characteristics of the big data including, but not limited to, velocity, volume, variety, and veracity, and we believe any auction mechanism design in the future should take the following factors into consideration: 1) generality (variety); 2) efficiency and scalability (velocity and volume); 3) truthfulness and verifiability (veracity). In this paper, we propose a privacy-preserving construction for auction mechanism design in the big data, which prevents adversaries from learning unnecessary information except those implied in the valid output of the auction. More specifically, we considered one of the most general form of the auction (to deal with the variety), and greatly improved the the efficiency and scalability by approximating the NP-hard problems and avoiding the design based on garbled circuits (to deal with velocity and volume), and finally prevented stakeholders from lying to each other for their own benefit (to deal with the veracity). We achieve these by introducing a novel privacy-preserving winner determination algorithm and a novel payment mechanism. Additionally, we further employ a blind signature scheme as a building block to let bidders verify the authenticity of their payment reported by the auctioneer. The comparison with peer work shows that we improve the asymptotic performance of peer works' overhead from the exponential growth to a linear growth and from linear growth to a logarithmic growth, which greatly improves the scalability.
\end{abstract}

%% file: intro_v1.tex
\section{Introduction}

Increasingly many decisions are made based on the data because of the rich information hidden behind it, and more and more data is being collected almost everywhere nowadays, which will soon lead us to the big data era. Among many `V's characterizing the big data, we focus on the 4`V's in this paper: variety, volume, velocity, and veracity. The starting point of this research is the observation that various auction mechanisms are adopted in different fields. Spectrum auction \cite{feng2013tahes,huang2013spring}, cellular networks \cite{dong2013ideal}, ad hoc networks \cite{li2013economic}, cloud computing \cite{lin2010dynamic}, cognitive radio networks \cite{chen2010auction,wang2010spectrum}, web advertisement \cite{borgs2007dynamics}, and smart grids \cite{wijaya2013matching} are good examples. However, the large and diverse pool of the information available for attackers in the big data has increased the privacy concerns \cite{laurila2012mobile,schadt2012changing,tene2012privacy}, and we present how to enable auctions in the big data context with 4Vs without privacy implications.

\subsection{Variety}

Different types of information is available from different sources for different parties in the big data, and the auctions may involve different types of goods. Existing solutions \cite{kolesnikov2009improved,
cachin1999efficient,brandt2005efficient,
suzuki2001efficient,stubblebine1999fair,
naor1999privacy} only deal with single-good auctions and thus lack general applicability in the big data context. To deal with such a variety in the big data, we target at a more general form of the auction than the simple ones which sell only one good at each auction -- Combinatorial Auction (CA hereafter). In a single-auctioneer CA, the auctioneer sells multiple heterogeneous goods simultaneously, and bidders bid on any combination of the goods instead of just one. Such auctions have been researched extensively recently \cite{cramton2006combinatorial,lehmann2002truth,
sandholm2001cabob,sandholm2002algorithm}, in part due to the generality of it, and in part due to growing applications in which combinatorial bidding is necessary \cite{fukuta2011toward,milgrom2000putting,kumar2006auction,
zaman2013combinatorial}. 

As further discussed in the following sections, the consideration of the combinatorial auction will bring great challenges to the auction design because of its inherent complexity.

\subsection{Velocity and Volume}

The velocity at which data is generated is at the different order of magnitude in the big data from the one in the traditional data, which pushed the volume of the processed data beyond PB, EB, and even ZB ($10^9$ TB) \cite{bigdata_growth}. The velocity and the volume in the big data brings great challenges into the realization of the privacy-preserving auction design in the following two aspects.

Firstly, an early work \cite{palmer2011development} relies on the secure multi-party computation using garbled circuits \cite{yao1986generate} and oblivious transfers \cite{rabin2005exchange} to solve the CA in a privacy-preserving manner. Such works protect the private information due to the powerful secure multi-party computation, but the circuit size grows very fast \textit{w.r.t.} the CA parameters (number of bidders, range of the bid value, maximum bid, number of goods) which leads to non-polynomial time computation time. Also, the oblivious transfer required for every gate in the circuit introduces a huge communication time as well. Therefore, the works based on garbled circuit are hardly applicable in the big data environment due to the inherent scalability and performance issue. Secondly, the combinatorial auction itself is a computationally hard problem. Even with assumptions which limit bidders' bidding behaviors (e.g., assuming \textit{single-minded} bidders \cite{lehmann2002truth,mu2008truthful}), CA typically requires to solve one or more NP-hard optimization problems, which leads to infeasible generic theoretical designs \cite{mackie1994generalized,varian1995economic}. Consequently, several works \cite{suzuki2003secure,yokoo2002secure} avoiding garbled circuit or oblivious transfer remains impractical because those solutions rely on the dynamic programming to calculate the optimum solution, which leads to a super-polynomial run time.

To address the scalability and performance issue to deal with the volume and velocity of the big data, we exclude the garbled circuits and oblivious transfers in our design, and further replace the exact optimization with the approximated one. This raises another challenge: traditional mechanism designs in CA guarantees truthful bidding to potentially maximize the social welfare based on the assumption that the goods are allocated optimally. Then, those mechanisms do not provide the same guarantee in our setting because we seek for the approximated result. Therefore, we cannot simply implement an existing approximation algorithm in a privacy-preserving manner, and we also need to improve existing mechanisms to preserve the truthfulness (defined later) of the auction.

\subsection{Veracity}

Data source in the big data is \textit{almost everywhere} in the world due to the proliferation of the data collection, and most of the data sources are not under strict quality control. Consequently, not all data in the big data era will be credible because of many reasons (\textit{e.g.,} machine factors: errors/inaccuracy/noises; human factors: moral hazard, mistake, misbehavior). In the CA, the veracity issue has an especially great impact because 1) the lying bidder may negatively affect the social welfare or auctioneer's total revenue; 2) and the winners' payments are calculated by the auctioneer, who is well motivated to report a higher fake price. Obliviously (\textit{i.e.,} without knowing the bid values or winners list) achieving a two-way verifiability against both untrusted bidders and auctioneers is another challenge in the privacy-preserving CA construction.

\subsection{Contributions}
The contribution of this work is prominent. This is the first paper to envision the privacy-preserving auction  mechanism in the upcoming big data age, which is designed based on the four main characteristics (variety, volume, velocity, and veracity) of the big data, and the contributions can also be summarized based on the 4V's: considering the variety of the big data, we explore privacy-preserving constructions for one of the most general auctions, CA; we have designed a scalable and efficient privacy-preserving algorithm to deal with the volume and velocity; and our design also provides two-way verifiability against malicious bidders and auctioneer to be robust to the veracity issue in the big data.\medskip

Note that our research does not explicitly work for the anonymity of the bidder or the auctioneer, but in fact our work is the last step of the anonymization. Our work complements the simple anonymization which replaces users' personally identifiable information (PII) with pseudo-random PIIs in the following sense. Such anonymization is vulnerable to various de-anonymization attacks \cite{danezis2013you,farenzena2010person} because published attributes can be fingerprinted or co-related with other datasets. By applying our on top of the sanitization, such de-anonymization becomes much more  challenging because the attributes of any tuple is protected as well.

\CUT{

\subsection{Main Challenges}
One of the main challenges in our privacy-preserving auction is to let the auctioneer determine the winners without knowing losers' selection of goods. This seems to be contradictory since a bidder's selection should be examined by the auctioneer to decide whether he is the loser. 

%However, we employ our past work \cite{jung2013data} as a building block to solve this non-trivial problem.

Next one is to preserve the truthfulness of the CA after replacing the exact optimization in the winner determination with an approximated one. 

%One mechanism is introduced in \cite{lehmann2002truth}. Although a bidder's utility cannot be increased by behaving dishonestly in \cite{lehmann2002truth}, an honest bidder's utility may be negative in some cases. We propose a new mechanism which guarantees the truthfulness and the non-negativeness of honest bidders' utility.

The last main challenge is to verify the payment reported by the auctioneer when the auction terminates. It is highly possible that an untrusted third party auctioneer may misbehave to achieve illegal monetary gain. Therefore, a winning bidder should be able to verify the payment without knowing other irrelevant information.

Besides, the confidentiality of the bid or the goods selection is somewhat obviously required in any privacy-preserving auction design and thus not highlighted, but the privacy is indeed preserved throughout our privacy-preserving CA construction.\medskip

The rest of the paper is organized as follows. More backgrounds of CA as well as related works are presented in Section \ref{section:background} to show why our work is viable compared to them, and the preliminaries are described in Section \ref{section:preliminary}. The privacy-preserving CA is modelled in Section \ref{section:model}, and the actual design is shown in Section \ref{section:construction}. We evaluate our construction in Section \ref{section:evaluation} and Section \ref{section:evaluation2} and finally conclude in Section \ref{section:conclusion}. 
}

%% file: background_v1.tex
\section{Preliminaries \& Related Work}\label{section:background}

\subsection{Backgrounds of Combinatorial Auction}
\begin{figure}[h]\label{fig:CAmodel}
\centering\vspace{-8pt}
\includegraphics[scale=0.25]{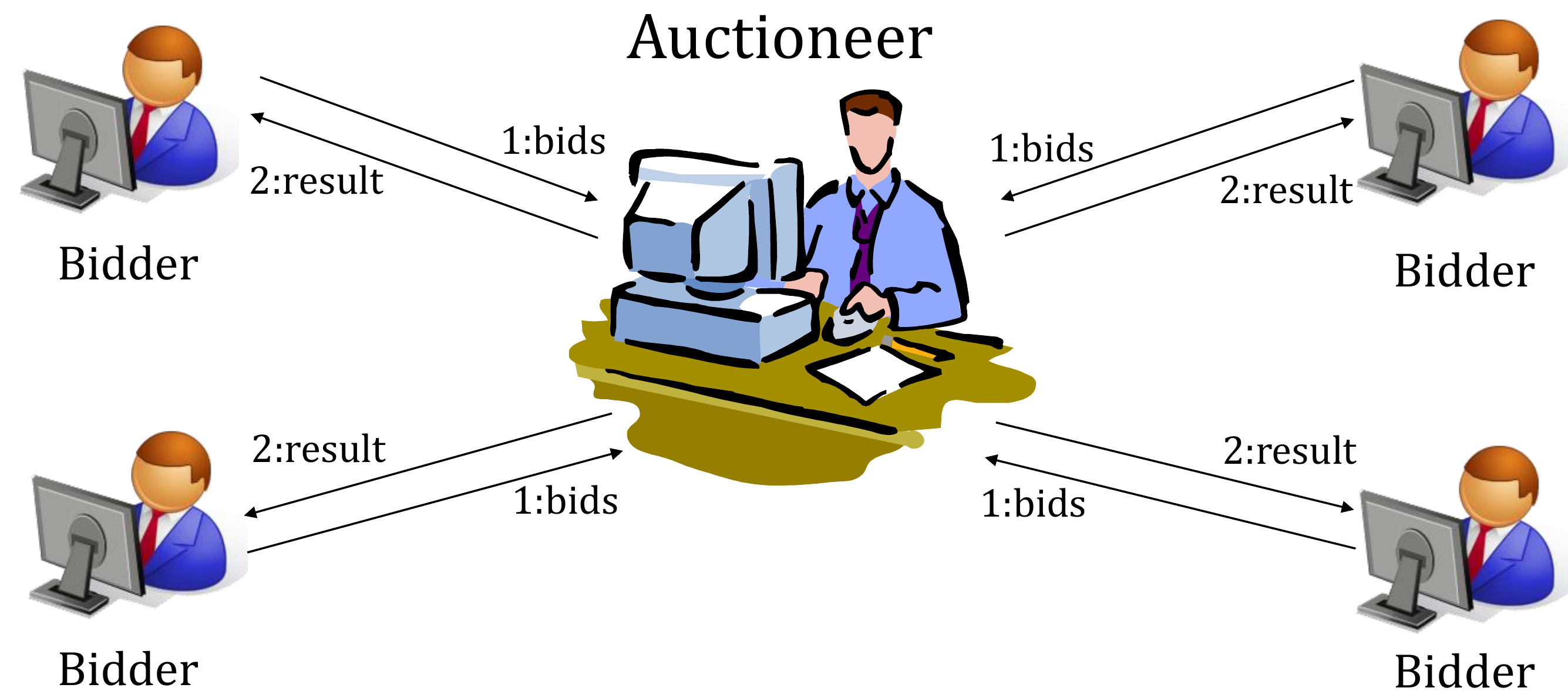}\vspace{-5pt}
\caption{One-stage, sealed-bid and single-sided Combinatorial Auction}\vspace{-5pt}
\end{figure}

Among various types of combinatorial auctions \cite{cramton2006combinatorial}\cite{rothkopf1983bidding}, we shall consider the most common type in this work, one-stage, sealed-bid and single-sided CA (Fig. \ref{fig:CAmodel}). In such auctions, each bidder places several bids, the auction terminates and the results are announced (one-stage); no information about other's bids is released prior to the auction termination (sealed-bid); and one auctioneer is selling several goods to multiple bidders (single-sided).

%\subsection{Bid Characteristic}
%Each buyer (henceforth bidder) may place several bids for each subset of the goods. Note that there exist the super-modularity and sub-modularity in the bids for each set, in that more goods in a set bring larger marginal gain and less marginal gain respectively. In some situations, a bidder may want to pay more for a whole set than the sum of what he is willing to pay for each part (super-modularity). This occurs when each part complements each others (e.g., two performance tickets with adjacent seats). In other situations, a bidder only wants to pay less for the whole than what he pays for the parts (sub-modularity). This is the case when, but not limited to, goods are similar and thus substitutable (e.g., two performance tickets with separate seats).

%\subsection{Auction Design}
Given such a CA, its mechanism design is composed of two parts. Firstly, winners of the auction are chosen based on their submitted bundles and bids \textbf{winner determination}, then each winner's payment is determined by some mechanism \textbf{payment determination}. Note that a winner's payment may not be equal to hid bid. 

\noindent \textbf{Winner Determination and Objective Function}

The standard goal of the design is to maximize the social welfare \cite{lehmann2002truth,dobzinski2005approximation,yokoo2004secure}, which is the sum of winners' reported bids on their allocated goods. An alternative goal is to maximize the auctioneer's revenue, which is the sum of winners' payments. Maximizing the revenue is closely related to the social welfare maximization, therefore we focus on how to maximize the social welfare. As aforementioned finding an allocation maximizing the social welfare is NP-hard \cite{rothkopf1998computationally}, and it has been shown that the optimal allocation can be approximated within a factor of $O(m^{\frac{1}{2}})$ but not to a factor of $O(m^{\frac{1}{2}-\epsilon})$ for any $\epsilon>0$ \cite{lehmann2002truth,sandholm2002algorithm}, where $m$ is the number of total goods. 

%The efficient allocation algorithm is presented in Section \ref{section:construction}. In a nutshell, each submitted bundle is sorted by the norm $\frac{b}{\sqrt{|S|}}$, where $b$ is the bid and $|S|$ is the number of goods in the bundle, and allocate the bundle if possible in the sorted order.

%\subsection{NP-Hardness of Winner Determination}
%Winners should be chosen so that the social efficiency is maximized, which is the sum of bidders' bids. However, this is an NP-hard problem as follows.
%
%\begin{theorem}
%Winner determination problem in the combinatorial auction is an NP-hard problem.
%\end{theorem}
%\begin{proof}
%The weighted maximum set packing problem (WMSPP), which requires to find pair-wise disjoint sets whose total weights is maximized, is known to be NP-hard \cite{chandra1999greedy,arkin1998local}. If we let each set be a bundle, weight be a bid, and the universe set be the set of all goods in the combinatorial auction, the WMSPP is reduced to the winner determination problem in a polynomial time. Therefore, the winner determination problem is an NP-hard problem.
%\end{proof}
%
%In fact, it is not only NP-hard, 

\noindent \textbf{Payment Determination and the Truthfulness}

Each bidder's bid may not truly reflect his valuation of the bundle. The payment is determined by all bidders' bids, and therefore bidders may try to report a fake valuation to decrease their payment or win a chance to win the auction.\vspace{-5pt}

\begin{definition}
An auction is truthful if reporting a true valuation is a weakly dominant strategy for every bidder, and utility of any honest bidder is non-negative. 
\end{definition}

\noindent That is, no bidder can increase his benefit by lying no matter other bidders lie or not.  Naturally, the payment mechanism determines whether the auction is truthful, and the one in the famous Generalized Vickrey Auction (GVA, \cite{mackie1994generalized}) guarantees the truthful auction, but determining one bidder's payment requires finding an optimum allocation without him, which is already shown to be NP-hard. Therefore, it is infeasible to implement the GVA in reality, and we study the truthfulness in conjunction with the aforementioned approximation.

A truthful mechanism for the approximated allocation is introduced in \cite{lehmann2002truth}. Let $L$ be the sorted list of bundles in the greedy allocation (sorted by bidders' norm $\frac{b}{\sqrt{|S|}}$). For any bundle $i$, denote the first bundle $j$ in $L$ which would have been allocated if $i$ were denied at first as \textbf{candidate} of $i$. Then, $i$'s payment is $
\frac{b'}{\sqrt{|S'|}}\cdot \sqrt{|S|}$
where $b'$ is the bid of the candidate bundle, $S'$ is the candidate bundle, and $S$ is the allocated bundle $i$. This payment guarantees the truthfulness of the auction, which is proved later in this paper  (Section \ref{section:evaluation}).

\subsection{Privacy-preserving Combinatorial Auctions}

Various approaches are proposed to achieve a private sealed-bid auction \cite{kolesnikov2009improved,cachin1999efficient,
brandt2005efficient,suzuki2001efficient,
stubblebine1999fair,naor1999privacy}, but much less attention is paid to the combinatorial auction. In general, recently proposed approaches for the secure multi-agent combinatorial auction can  be divided into two classes: first class based on Secure Multi-party Computation (SMC) and the other class based on Homomorphic Encryption (HE).

To the best of our knowledge, \cite{palmer2011development,pan2012using,yokoo2002secure,suzuki2003secure} 
are the only works solving CA in a privacy-preserving manner. \cite{palmer2011development} solves it by leveraging SMC, but as shown in Palmer's implementation, the solution based on SMC does not scale well because the circuit needs to implement all if-else branches in it in order to accept arbitrary input. Besides, \cite{yokoo2002secure,suzuki2003secure} designed a secure multi-agent dynamic programming based on HE, which is in turn used to design the privacy-preserving winner determination in CA. Pan \textit{et al.} \cite{pan2012using} also designed a combinatorial auction based on HE. However, these all target at solving the optimum solution for the winner determination problem, and their protocols cannot be run in polynomial time due to the inherent hardness of the problem.

Besides, our work has one more advantage: our auction scheme is the only one which presents a privacy-preserving payment determination mechanism to guarantee the truthfulness of the auction, while all aforementioned works only solve the winner determination problem in the combinatorial auction.

%% file: model_v1.tex
\section{Privacy-preserving Combinatorial Auction Model}\label{section:model}

\subsection{Auction Model}
A set of $m$ goods $G=\{g_1,\cdots,g_m\}$ are auctioned to $n$ bidders $\{\bidder_1,\cdots,\bidder_n\}$ in the CA, and the \textit{combinatorial} characteristic comes from each bidder $\bidder_i$'s bid and valuation on a \textit{bundle} $S_i\subseteq G$ instead of each good. $\bidder_i$ proposes his bid $b_i(S_i)$ (i.e., maximum willingness to pay) on the  bundle $S_i$, and the bid might be different from his true valuation $v_i(S_i)$ if he wishes to lie. A set $W$ of winners are chosen by the auctioneer as follows:

{\footnotesize
\begin{displaymath}
W=\text{argmax}_X{\sum_{\bidder_i\in X}{b_i(S_i)}} ~~s.t.~~ \bigcap_{\bidder_i\in X}{S_i}=\emptyset 
\end{displaymath}\vspace{-6pt}}

\noindent i.e., a set of conflict-free bidders whose social welfare is maximized, and the corresponding allocation set $A^*$ is $A=\bigcup_{\bidder_i\in W}S_i$. After the winners are chosen, each winner $\bidder_i$'s payment $p_i$ is determined 
by the auction mechanism based on all bidders bids. Then, we assume a quasilinear utility for every bidder as follows:

{\footnotesize
\begin{displaymath}
u_i=\begin{cases}
v_i(S_i)-p_i & \bidder_i\text{ is a winner}\\
0 & \text{Otherwise}
\end{cases}
\end{displaymath}\vspace{-8pt}}

We assume bidders are \textbf{single-minded}. That is, each of them cares only about one specific set of goods, and if they do not get the desired set, their valuation on the result is 0. Formally, for any $\bidder_i$'s desired set $S_i$, $\bidder_i$'s valuation on a set $S'$ is $v_i$ if and only if $S_i\subseteq s'$. This assumption is equivalent to the restriction that each bidder is limited to one bid only. Therefore, we simplify the notation $v_i(S_i)$ as $v_i$ and $b_i(S_i)$ as $b_i$ hereafter.\vspace{-4pt}

\begin{table}[h]
\centering\vspace{-8pt}
\caption{Frequently Used Notations}\vspace{-7pt}
\begin{tabular}{r|l}\hline\hline
Auctioneer & $\auctioneer$ \\
$i$-th bidder & $\bidder_i$\\
$\bidder_i$'s bundle & $S_i$\\
$\bidder_i$'s bid& $b_i$\\
\hline\hline
\end{tabular}\label{table:notation}\vspace{-12pt}
\end{table}

\subsection{Adversarial Model}
When the auction terminates, $\auctioneer$ is supposed to know only the winners, their bundles and corresponding payments. Each bidder $\bidder_i$ only learns whether he is the winner when the auction terminates, and he is informed of the payment if he is chosen as a winner. He does not learn anything about others' bids or bundles except what implied in his payment (which is very limited information).

The auctioneer is assumed to be \textbf{curious}, \textbf{malicious} and \textbf{ignorant}. He is interested in bidders' bids and bundles to improve his business (called ``curious''). For example, he may try to infer bidders' preferences and rivalry relationship based on the bids and the bundles. The auctioneer may also report a fake payment to the winners to illegally increase his revenue (called ``malicious''), but he is not aware of bidders' side information such as distribution of bid values or bidders' preferences on goods (called ``ignorant'').

Bidders are assumed to be \textbf{selfish}, \textbf{curious} and \textbf{non}-\textbf{cooperative}. Their objective is to maximize their own utilities, and bidders will report fake valuations if the utility is increased by doing so (called ``selfish''). On the other hand, bidders are interested in others' bids and bundles to improve the decision making (called ``curious''). However, they will not collude with other bidders or the auctioneer (called ``non-cooperative'').

\subsection{Privacy Definitions}
\begin{definition}
Given all the communication strings $\mathcal{C}$ during the auction and the output of the auction \textsf{Output}, an adversary's advantage over the loser $\bidder_i$'s bid $b_i$ is defined as
\end{definition}\vspace{-12pt}

{\footnotesize
\begin{displaymath}
adv_{b_i}=\text{Pr[$b_i$}|\mathcal{C},\textsf{Output}\leftarrow \mathcal{A}_{our}(1^\kappa)]-\text{Pr[$b_i$}|\textsf{Output}\leftarrow \mathcal{A}_{black}]
\end{displaymath}\vspace{-15pt}}

\noindent \textit{where} Pr[$b_i$] \textit{is the probability that a correct $b_i$ is inferred, $\mathcal{A}_{our}$ is our algorithms in the protocol, $\kappa$ is the security parameter of them, and $\mathcal{A}_{black}$ is a perfectly secure black-box algorithm which only returns the final results without any side information.}

We focus on the confidentiality of auction losers' bids in this paper because winners' bids can be learned from the valid outputs of the auctions anyway (\textit{e.g.,} claimed bundle and the payments).

\begin{definition}
Given all the communication strings $\mathcal{C}$ during the auction and the output of the auction \textsf{Output}, an adversary's advantage over the bidder $\bidder_i$'s bundle $S_i$ is defined as
\end{definition}\vspace{-12pt}

{\footnotesize
\begin{displaymath}
adv_{S_i}=\text{Pr[$S_i$}|\mathcal{C},\textsf{Output}\leftarrow \mathcal{A}_{our}(1^\kappa)]-\text{Pr[$S_i$}|\textsf{Output}\leftarrow \mathcal{A}_{black}]
\end{displaymath}\vspace{-15pt}}

\noindent \textit{where} Pr[$S_i$] \textit{is the probability that any information about $S_i$ is inferred, and other notations are same as in Definition 2.}\smallskip

Informally, these advantages measure how much side information an adversary gains during our privacy-preserving auction by measuring the increased probabilities. In other words, they reflect how much side information is disclosed other than what is derivable from the valid auction output.

%% file: preliminary_v2.tex
\section{Building Blocks}\label{section:preliminary}
Before we introduce our auction design, we first introduce several building blocks as well as preliminary techniques used in our design.

\subsection{Homomorphic Encryption}
Homomorphic encryption allows specific computations to be directly carried on ciphertexts while preserving their decryptability. El Gamal encryption \cite{elgamal1985public}, Goldwasser–Micali cryptosystem \cite{goldwasser1982probabilistic}, Benaloh cryptosystem \cite{benaloh1994dense} and Paillier cryptosystem \cite{paillier1999public} are good examples, and we employ the Paillier cryptosystem to implement a one-way privacy-preserving scalar product for efficient our winner determination. The Paillier cryptosystem  as well as its homomorphic property is shown below:

{\footnotesize
\begin{table}[h]
\centering
\begin{tabular}{|l|}
\hline
~~~~~~~~~~~~~~~~~~~~~~~\textbf{Paillier Cryptosystem}\\
\textbf{System parameter}: two prime numbers $p,q$.\\
\textbf{Public key}: modulus $n=pq$ and a random number $g\in \mathbb{Z}^*_{n^2}$\\
\textbf{Private key}: $\lambda=LCM(p-1,q-1)$\\
\textbf{Encryption}: $c=E(m,r)=g^{m+nr}~~\mod n^2$\\ ~~~~~~~~~~~~~~~~where $r\in \mathbb{Z}^*_{n}$ is a random number.\\
\textbf{Decryption}: $m=D(c)=\frac{L(c^\lambda~~\mod n^2)}{L(g^\lambda~~\mod n^2)}~~\mod n$\\
~~~~~~~~~~~~~~~~where $L(x)=\frac{x-1}{n}~~\mod n$\\
\textbf{Self-blinding}: $E(m,r)\cdot g^{nr'}=E(m,r+r')$\\\hline
\end{tabular}
\end{table}

\begin{displaymath}
\left.
\begin{aligned}
E(m_1,r_1)\cdot E(m_2,r_2) &= E(m_1+m_2,r_1+r_2)\\
E(m_1)\cdot g^{m_2} &= E(m_1+m_2,r_1)
\end{aligned}\right\}\text{Addition}
\end{displaymath}
\begin{displaymath}
E(m_1,r_1)^{m_2}=E(m_1\cdot m_2,r_1\cdot m_2)-\text{Multiplication}
\end{displaymath}

}

\subsection{Digital Signature}

  %A digital signature is a demonstration of the intactness of a digital value. A digital value along with its valid signature enables the recipient to verify that the value is not altered after the signature is generated. 
  
  In our work, we employ a signer who is involved only to generate a signature of each bidder's value. Many works can be considered \cite{goldwasser1988digital,merkle1990certified,
  nyberg1993new}, but we use the blinded Nyberg-Rueppel scheme in \cite{camenisch1995blind} (Table), which is a blind signature scheme. In a blind signature scheme, the signer can generate a signature of a value $m$ without `seeing' it. In our work, we use the mechanism design to guarantee the truthfulness of the bidding, and use a blind signature scheme to verify the payment is calculated correctly. Since the authenticity of the bids are guaranteed by the truthful mechanism, we do not need the signer to verify the authenticity of them, therefore a blind signature scheme suffices. Notably, the recipient can recover the message from the signature in this scheme. 
For the simplicity, we denote the signature of $m$ as $\text{Sig}(m)$ hereafter.

\floatname{algorithm}{Scheme}
\begin{algorithm}
\caption{Blinded Nyberg-Rueppel Scheme}
\textbf{System parameter}: a multiplicative group $\mathbb{G}\subset\mathbb{Z}_p^*$ of prime order $q$ and its generator $g$, where $q$ is a prime factor of prime number $p$.

\textbf{Key Generation}: signer picks a random number $x\in\mathbb{Z}_q$. Then, he keeps $x$ secret and publishes the public parameters as $g,g^x~(\text{mod}~ p)$.

\textbf{Signing}: signer blindly signs signee's message $m$.

\begin{algorithmic}[1]
\STATE The signer randomly selects $\hat{k}\in\mathbb{Z}_q$ and sends $\hat{r}=g^{\hat{k}}~(\text{mod}~ p)$ to the signee.

\STATE The signee randomly selects $\alpha\in\mathbb{Z}_q,\beta\in\mathbb{Z}_q^*$, computes $r=mg^\alpha~(\text{mod}~ p)$ and $\hat{m}=r\beta^{-1}~(\text{mod}~ q)$ until $\hat{m}\in\mathbb{Z}_q^*$. Then, he sends $\hat{m}$ to the signer.

\STATE The signer computes $\hat{s}=\hat{m}x+\hat{k}~(\text{mod}~ q)$ and sends $\hat{s}$ to the signee.

\STATE The signee computes $s=\hat{s}\beta+\alpha~(\text{mod}~ q)$, and the pair $(r,s)$ is the Nyberg-Rueppel signature for $m$. 
\end{algorithmic}
\textbf{Verification}: check whether $m=g^{-s}y^rr~(\text{mod}~ p)$.
\end{algorithm}

 Not all blind signature schemes can be used. The ones using homomorphic encryption is often subject to this problem as follows:

{\footnotesize\vspace{-11pt}
\begin{displaymath}
\begin{split}
\text{Sig}(v_1)^{v_2}
=E_{EK}(v_1)^{v_2}=E_{EK}(v_1 v_2)=\text{Sig}(v_1 v_2)
\end{split}
\end{displaymath}\vspace{-13pt}}

\noindent where `Sig' stands for the signature. Then, the signature of value $v_1v_2$ is illegally generated from the $\text{Sig}(v_1)$. In addition, the DSA signature scheme mentioned in  \cite{camenisch1995blind} is also subject to forgery. Given a signature $\text{Sig}(m)=(s,r)$ for $m$, an attacker can create a fake signature $\text{Sig}(km)=(ks,\frac{r}{k})$.

%Secondly, we cannot use a blind signature scheme or zero-knowledge proof (ZKP) since the purpose of those two is to prove the ownership of the value, but we need to authenticate whether bidders are using fake values throughout the auction, therefore the third-party needs to `look at' the value to examine the authenticity. Therefore, we use any simple non-homomorphic public key encryption scheme to sign the valid values. 

\subsection{Fixed-point Representation}\label{section:integer_real}

Due to the cryptographic operations in our work, all the computations and operations in our scheme are closed in integer groups, and therefore our scheme cannot be directly applied when the numeric type is real number. 

%However, we can exploit the homomorphism\footnote{$C(m_1)m_2=C(m_1m_2)$, $E_\HPK(m_1)E_\HPK(m_2)=E_\HPK(m_1+m_2)$, $E_\HPK(m_1)^{m_2}=E_\HPK(m_1m_2)$} of our scheme to use integers to represent real numbers.

Floating point representation ($[a]=\{m,e\}$ where $a\approx 1.m\cdot 2^e$) is one way to use integers to represent real numbers, but it is hard to implement the arithmetic homomorphic operations (addition, multiplication) on floating point representations. Instead, we use the fixed point representation \cite{nikolaenko2013privacy} to represent real numbers due to its simplicity when applying elementary arithmetic operations to it. Given a real number $a$, its fixed point representation is given by $[x]=\lfloor x\cdot 2^e \rfloor$ for a fixed $e$, and all the arithmetic operations reduce to the integer versions as follows:
\begin{compactitem}
\item Addition/Subtraction: $[x\pm y]=[x]\pm [y]$

\item Multiplication/Division : $[x\cdot y^{\pm 1}]=[x]\cdot [y]^{\pm 1}\cdot 2^{\mp e}$
\end{compactitem}
where $x\cdot 2^{-1}$ stands for $x$ divided by 2. Then, our protocols in the following sections can be trivially extended to real-number domain.

%% file: construction_v2.tex
\section{Designing Auction Mechanism For Big Data}\label{section:construction}

Before the auction proceeds, all bidders are asked to blindly sign their $\psi_i$ and $S_i$\footnote{$S_i$ is a set, but it can be mapped to an integer with any 1-to-1 mapping (\textit{e.g.,} hash).} via a third-party signer $\mathcal{T}$. Since the signature is blindly signed, $\mathcal{T}$ learns nothing. These signatures will be used to verify the authenticity of the bids later.

\subsection{Privacy-preserving Winner Determination}

\floatname{algorithm}{Algorithm}
\begin{algorithm}
\caption{Greedy Winner Determination}
\begin{algorithmic}[1]
\STATE $A:=\emptyset, W:=\emptyset$. For each $\bidder_i$, computes $\psi_i=\frac{b_i}{\sqrt{
|S_i|}}$.

\STATE Sort the instances in the non-increasing order of norm $\psi_i$. Denote the sorted list as $L$.

\STATE For each $\bidder_i\in L$ (in the sorted order), check whether $A\cap S_i=\emptyset$. If true, $A:=A\cup S_i, W:=W\cup\bidder_i$.

\STATE $A^*:=A$. Announce $W$ as the winners. Finally allocated goods are $A^*$.
\end{algorithmic}
\end{algorithm}

The above approximation algorithm for the winner determination guarantees an approximation ratio of at least $O(\sqrt{m})$ \cite{lehmann2002truth}, where $m$ is the number of total goods, and this has been proved to be the best approximation ratio that can be achieved \cite{lehmann2002truth,sandholm2002algorithm}. To guarantee each bidder's privacy, we cannot explicitly perform the sorting (Step 2.) because the order of all bidders' norms reveals excessive side information about the losers' $b_i$ or $S_i$ even if the norm $\psi_i$ does not directly reveal either one. Therefore, we unravel Step 2. and Step 3. as follows. Firstly, among the \textit{encrypted} bundles that can be allocated (\textit{i.e.,} no overlap with already-allocated goods),  find out the one whose corresponding norm $\psi_i$ is the maximum (without revealing $\psi_i$'s value). Then, find out the winner who owns the bundle (up to previous step, every one was anonymous). Finally, update $A$ correspondingly and keeps looking for the next feasible bundle with maximum $\psi_i$. Note that the IDs are already anonymized either via sanitization or anonymized network such as Tor \cite{dingledine2004tor}, therefore only the winners' identities are revealed to the auctioneer. 

To further describe the privacy-preserving unraveled greedy algorithm more easily, we first elaborate a sub-procedure in it: feasibility evaluation.\smallskip

\noindent \textbf{Feasibility Evaluation}

Given a bundle $S_i$, whether it is feasible (\textit{i.e.,} does not overlap with already-allocated goods) must be evaluated in a privacy-preserving manner in order to keep the confidentiality of bids or bundles. Firstly, we use an $m$-dimension binary vector $\textbf{A}$ represent the allocation status of all goods (\textit{i.e.,} the goods in $A$), where the $k$-th bit $a_k=1$ if the $k$-th good $g_k$ is allocated already and 0 otherwise. Similarly, we use another vector $\textbf{S}_i$ to represent $\bidder_i$'s bundle $S_i$, where $\textbf{S}_i$'s $k$-th bit $s_{i,k}=1$ if $g_k\in S_i$ and 0 otherwise.

Then, $A\cap S_i=\emptyset$ if and only if

{\footnotesize
\begin{displaymath}
\textbf{A}\cdot \textbf{S}_i = 0
\Leftrightarrow\sum_{k=1}^{m}a_ks_{i,k}=0
\end{displaymath}\vspace{-5pt}}

If the scalar product is $\theta$, that means $\bidder_i$'s bundle $S_i$ includes $\theta$ already-allocated goods. In order to keep $\theta$ and $\{s_{i,k}\}$ secret from the auctioneer, and to keep $\theta$ and $\{a_k\}$ secret from $\bidder_i$, we propose the following privacy-preserving scalar product to let the auctioneer learn whether the above sum is equal to 0.

\begin{algorithm}[h]\label{alg:scalar}
\caption{Privacy-preserving Scalar Product}
\begin{algorithmic}[1]

\STATE $Auc$ picks a pair of Paillier cryptosystem key: $PK'_{Auc}=(n,g)$, $SK'_{Auc}=\lambda$ (Section \ref{section:preliminary}).

\STATE $Auc$ encrypts every bit $a_k$ homomorphically and sends its ciphertext $E_{Auc}(a_k)$ to the bidder $\bidder_i$ whose bundle is being checked.

\STATE Upon receiving $m$ ciphertexts, $\bidder_i$ first picks a random number $\delta_i\in\zgroup_n$ and performs following operations:

{\footnotesize\vspace{-4pt}
\begin{displaymath}
\forall k: c_k=E_{Auc}(a_k)^{\delta_i s_{i,k}}=E_{Auc}(\delta_i a_ks_{i,k})
\end{displaymath}\vspace{-12pt}}

Then, he computes the following and sends to the auctioneer:

{\footnotesize\vspace{-4pt}
\begin{displaymath}
c = \prod_{k=1}^m c_k = E_{Auc}(\delta_i\sum_{k=1}^ma_ks_{i,k})
\end{displaymath}\vspace{-7pt}}

\STATE The auctioneer decrypts the received ciphertext using his secret key, which is the scalar product $\delta_i\cdot \textbf{A}\cdot \textbf{S}_i$.
\end{algorithmic}
\end{algorithm}

%The auctioneer creates and controls one virtual aggregater and $m$ virtual participants, and the bidder $\bidder_i$ creates and controls $m$ virtual participants. Then, the virtual aggregater and $2m$ virtual participants are engaged in the Participants Only MPEP to compute the above product, which is a $2m$-variable polynomial. Both the auctioneer and the bidder $\bidder_i$ receive the evaluation result and know whether the $\bidder_i$ is one of the winner (if the outcome is 1), and $\bidder_i$ reports his bundle to the auctioneer if he is the winner and remains silent otherwise.

If $\delta_i \cdot \textbf{A}\cdot \textbf{S}_i=0$, $Auc$ learns $S_i$ is feasible, and if $S_i$ is not feasible, the outcome is $\delta_i\theta$ which is indistinguishable to a random number in $\zgroup_n$ from the auctioneer's perspective because of the randomizer $\delta_i$.

Combining the above feasibility evaluation, we are ready to present our privacy-preserving winner determination algorithm which works as a black-box algorithm outputting the winning bundles and the winners only (Algorithm \ref{alg:winnerdetermination}). Essentially, the outcome of 3-b is 0 if and only if $\bidder_i$'s $\psi_i$ is equal to the $Auc$'s guess, and $c$ at 3-c is equal to 0 if and only if $\bidder_i$'s $S_i$ is feasible at the current allocation $A$. Then, the final outcome at 3-e  is equal to 0 if and only if $\bidder_i$'s $\psi_i$ is the maximum among all the remaining bidders and his bundle is also feasible. 

Besides this, the algorithm deserves further clarifications at the places terms or phrases are marked bold. Firstly, the way $Auc$ guesses the maximum norm at step 3 is critical for the performance. Given the range of the possible values for the norms, $Auc$ performs a binary search until finding a value $\psi^*$ such that the final outcome at 3-e yields 0 at $\psi^*$ but not at $\psi^*+1$. If such values are discovered, the next binary search can be started from $\psi^*$. Secondly, given the outcome yielding 0 at 3-e, $Auc$ must find out the winner first because every bidder is
anonymous yet up to this point. This can be done by declaring the anonymous ID of the winner and asking him to reveal his $\psi_i,S_i,\text{Sig}(\psi_i)$, and $\text{Sig}(S_i)$ to auctioneer for the confirmation. Since everything was encrypted under $Auc$'s keys, no bidders gain any information about the $\psi^*$. On the other hand, because the entire protocol is conducted in an anonymized network, declaring the anonymous ID does not breach winner privacy (anonymous ID is often a one time identity). If $psi_i=\psi^*$, $Auc$ learns the bidder is the winner, and marks his goods as allocated in $A$ (the authenticity of $\psi_i$ and $S_i$ can be verified with the signatures). Finally, $Auc$ learns no more update is possible when he finds out the binary search is terminated but no one yielded 0 at 3-e.

\begin{algorithm}[h]\label{alg:winnerdetermination}
\caption{Privacy-preserving Winner Determination}
\begin{algorithmic}[1]
\STATE $A:=\emptyset, W:=\emptyset, B=\{\bidder_i\}_i$. Every $\bidder_i$ computes $\psi_i=\frac{b_i}{\sqrt{
|S_i|}}$ individually.

\STATE $Auc$ picks a pair of Paillier cryptosystem key: $PK=(n,g)$, $SK=\lambda$ (Section \ref{section:preliminary}), and publishes $PK$.

\STATE $Auc$ \textbf{guesses} the maximum $\psi$ value $\psi^*$, and checks whether there exists a bundle with $\psi_i=\psi^*$ that can be allocated by performing the following procedure with every bidder $\bidder_i\in B$.\smallskip

3-a: $Auc$ sends $E_{Auc}(\psi^*)$ (ciphertext of $\psi^*$) to $\bidder_i$.

3-b: $\bidder_i$ picks a random number $\delta_i'\in\zgroup_n$, then calculates: $$\Big(E_{Auc}(\psi^*)\cdot E_{Auc}(-\psi_i)\Big)^{\delta_i'}=E_{Auc}\Big(\delta_i'(\psi^*-\psi_i)\Big)$$

3-c: $Auc$ sends out encrypted $\{a_k\}$'s to $\bidder_i$, and $\bidder_i$ calculates $c=E_{Auc}(\delta_i\sum a_ks_{i,k})$ as in the aforementioned scalar product calculation (Algorithm \ref{alg:scalar}).

3-d: $\bidder_i$ sends the following to $Auc$: $$E_{Auc}\Big(\delta_i'(\psi^*-\psi_i)\Big)\cdot E_{Auc}\Big(\delta_i\sum a_ks_{i,k}\Big)$$ $$=E_{Auc}\Big(\delta_i'(\psi^*-\psi_i)+\delta_i\sum_{k=1}^m a_ks_{i,k}\Big)$$

3-e: $Auc$ decrypts it to see whether it is equal to 0.

\STATE Step 3 is repeated with different $\psi^*$ to find out the maximum $\psi^*$ yielding 0 in 3-e. If an anonymous bidder's outcome is discovered to yield 0 in 3-e, $Auc$ \textbf{finds out the winner}, mark the corresponding goods as allocated in $A$, and add $\bidder_i$ to $W$. Then, repeat 3. again with updated sets. This is repeated until \textbf{no more update is possible}.

\STATE Set $A^*=A$. Then, $Auc$ learns $W$ is the set of winners, and $A^*$ is the finalized allocation. Then, he proceeds to payment determination.
\end{algorithmic}
\end{algorithm}

\subsection{Privacy-preserving Verifiable Payment Determination}

In the truthful auction mechanism mentioned in Section \ref{section:background}, the auctioneer determines a winner $\bidder_i$'s payment as follows. Among the bidders  whose bundle would have been allocated if $\bidder_i$ were not the winner (i.e., the candidate of $\bidder_i$), the auctioneer finds out the one with the maximum $\psi$ (say $\psi_j$ of bidder $\bidder_j$). Then, $\bidder_i$'s payment is $
p_i=\frac{b_j}{\sqrt{|S_j|}}\sqrt{|S_i|}$.

Three parties are engaged here: auctioneer $Auc$, winner $\bidder_i$ and $\bidder_i$'s candidate $\bidder_j$. The auctioneer $Auc$ needs to know $p_i$ without knowing $\bidder_j$'s bundle or bid; the winner $\bidder_i$ needs to know $p_i$ without knowing $\bidder_j$'s bundle or bid, and he should not even know who is the $\bidder_j$; and finally, the bidder $\bidder_j$ does not need to know anything from this whole process. Furthermore, both the auctioneer and the winner should be able to verify the payment.  We present the privacy-preserving verifiable payment determination (Algorithm \ref{alg:payment}) which fulfills above requirements as follows. 

\begin{algorithm}[h]
\label{alg:payment}
\caption{$\bidder_i$'s Verifiable Payment Determination}
\begin{algorithmic}[1]

\STATE The auctioneer $Auc$ excludes $B_i$ from $B$, and finds out the winner with $(A^*-S_i)$ by following the same procedure as the winner determination, where $A^*$ is the finally sold goods. If $\bidder_j$ is the winner, then he is the candidate of $\bidder_i$. Different from the original winner determination, $\bidder_j$ only reveals $\psi_j=\frac{b_j}{\sqrt{|S_j|}}$ and $\text{Sig}(\psi_j)$ to the auctioneer for the confirmation.

\STATE If a candidate is found, $Auc$ calculates $p_i=\psi_j\sqrt{|S_i|}$ and sends $p_i$ as well as the $\text{Sig}(\psi_j)$ to $\bidder_i$. Otherwise, $Auc$ sets $p_i$ as the reserve price (\textit{e.g.,} pre-defined minimum price) and informs $\bidder_i$ that his payment is the reserve price.

\STATE If the payment is not the reserve price, $\bidder_i$ recovers $\psi_j$ from $\text{Sig}(\psi_j)$, and verifies whether $p_i= \psi_j\sqrt{|S_i|}$. If they are not equal toe each other, he learns that the payment is incorrect.
\end{algorithmic}
\end{algorithm}

Since $Auc$ uses aforementioned privacy-preserving feasibility evaluation, he does not learn about $\bidder_j$'s bundle, and therefore he does not learn $b_j$ from $\psi_j=\frac{b_j}{\sqrt{|S_j|}}$. The winner $\bidder_i$ does not learn about $b_j$ due to the same reason, and he  also does not know who is $\bidder_j$ since he does not even communicate with $\bidder_j$. On the other hand, owing to the signature $\text{Sig}(\psi_j)$ generated by $\signer$,  $Auc$ is convinced that $\bidder_j$ did not report a fake lower $\hat{n}_j$ to harm $Auc$'s business, and the winner $\bidder_i$ believes $Auc$ did not tell a fake higher $p_i$ to illegally increase $Auc$'s revenue.

%% file: evaluation_v1.tex
\section{Theoretical Properties of Our Protocols}\label{section:evaluation}

\subsection{Truthfulness and the Payment Mechanism}

The winner $\bidder_i$'s payment is determined by his candidate $\bidder_j$ whose bundle would be allocated if $\bidder_i$ were not. His payment is $p_i=\frac{b_j}{\sqrt{|S_j|}}\cdot S_i$.

\begin{theorem}
Any honest bidder's utility is non-negative.
\end{theorem}
\begin{proof}
$v_i=b_i$ for a honest bidder $\bidder_i$. If $\bidder_i$ is not a winner, his utility is $0$. Otherwise, his utility is

{\footnotesize\vspace{-6pt}
\begin{displaymath}
u_i=v_i-p_i=b_i-\frac{b_j}{\sqrt{|S_j|}}\cdot \sqrt{|S_i|}
\end{displaymath}\vspace{-7pt}}

\noindent Since $\bidder_j$ is behind $\bidder_i$ in the sorted list $L$, $\bidder_i$'s norm is greater than $\bidder_j$'s one. Then, 

{\footnotesize\vspace{-6pt}
\begin{displaymath}
\frac{b_i}{\sqrt{|S_i|}}\geq \frac{b_j}{\sqrt{|S_j|}}\Rightarrow b_i \geq \frac{b_j}{\sqrt{|S_j|}}\cdot |S_i|\Rightarrow u_i\geq 0
\end{displaymath}\vspace{-7pt}}

\noindent Therefore, any honest bidder's utility is non-negative.
\end{proof}

\begin{theorem}
Any bidder's utility is not increased when he bids dishonestly.
\end{theorem}
\begin{proof}
Any bidder $\bidder_i$'s utility is

{\footnotesize\vspace{-4pt}
\begin{displaymath}
u_i=\begin{cases}
v_i-\frac{b_j}{\sqrt{|S_j|}}\cdot \sqrt{|S_i|} & \bidder_i \text{ is a winner}\\
0 & \text{Otherwise}
\end{cases}
\end{displaymath}\vspace{-6pt}}

\noindent We discuss two different cases of $\bidder_i$'s valuation $v_i$. Again, $\bidder_j$ is $\bidder_i$'s candidate.\smallskip

\noindent \textbf{Case 1}: {\footnotesize $v_i<\frac{b_j}{\sqrt{|S_j|}}\cdot \sqrt{|S_i|}$}

In this case, if $\bidder_i$ bids honestly, $\bidder_j$ becomes the winner, and $\bidder_i$ gains 0 as his utility. This is same when he reports a lower bid than $v_i$, or he reports a higher bid than $v_i$ but not high enough to beat $\bidder_j$ and becomes a winner. However, if he bids dishonestly by reporting a higher bid than $v_i$ and becomes the winner, his utility is

{\footnotesize\vspace{-5pt}
\begin{displaymath}
u_i=v_i-\frac{b_j}{\sqrt{|S_j|}}\cdot \sqrt{|S_i|}
\end{displaymath}\vspace{-8pt}}

\noindent Since $\bidder_i$ is not a winner if $b_i=v_i$, we have:

{\footnotesize\vspace{-5pt}
\begin{displaymath}
\frac{v_i}{\sqrt{|S_i|}}<\frac{b_j}{\sqrt{|S_j|}}\Rightarrow v_i<\frac{b_j}{\sqrt{|S_j|}}\cdot \sqrt{|S_i|} \Rightarrow u_i < 0
\end{displaymath}\vspace{-8pt}}

\noindent Therefore, bidders do not gain benefit by lying in this case.\smallskip

\noindent \textbf{Case 2}: {\footnotesize $v_i\geq \frac{b_j}{\sqrt{|S_j|}}\cdot \sqrt{|S_i|}$}

In this case, if $\bidder_i$ bids honestly, he is the winner and achieves a non-negative utility (Theorem 2). This is same when he reports a higher bid than $v_i$, or he reports a lower bid than $v_i$ but not low enough to lose the auction. However, if he bids dishonestly by reporting a lower bid than $v_i$ and loses the auction, his utility becomes 0. Therefore, bidders do not gain benefit by lying in this cases.
\smallskip

In conclusion, bidders do not increase their utility by reporting a fake valuation as bid in this payment mechanism.
\end{proof}

Combining the Theorem 1 and 2, we can conclude that our payment mechanism guarantees  the truthfulness of the auction.

\CUT{
\subsection{Intractability Assumption}

We assume the discrete logarithm problem is intractable in our work as in similar cryptographic works (\cite{jung2013privacy,paillier1999public,li2013search,elgamal1985public,zhang2013verifiable}). We prove that no one is able to achieve privacy-preserving winner determination without this assumption.

%Kushilevitz \cite{kushelvitz1992privacy} has proved that a two-party function $f$ is privately computable if and only if $f$'s function matrix is \textit{decomposable}. 
%
%Given two parties' input sets $C$ and $D$, the $f$'s function matrix $M_f$ is a $|C|\times |D|$ matrix which lists all function output given an input $(x,y)\in C\times D$.
%
%\begin{definition}
%A matrix is rows- (columns-) decomposable if there is a partitioning or rows (columns) so that there is no pair from two partitions which share an identical entry at the same column (row).
%\end{definition}
%
%\begin{definition}
%A matrix is decomposable if it is monochromatic (i.e., having identical elements at all entries), or it is rows- or columns-decomposable to a decomposable submatrices.
%\end{definition}

%No sufficient condition to the private computability for $n$-party functions is found yet, but 

Chor and Kushilevitz \cite{chor1989zero} gave a necessary condition for the unconditionally private computability (without intractability assumption) of $n$-party functions, and it is developed to the following lemma by Brandt \textit{et al.} \cite{brandt2008existence}.

\begin{lemma}
Given $\bar{x}=(x_1,\cdots,x_{n-1})$, $\bar{y}=(y_1,\cdots,y_{n-1})$, $x_{n}$ and $y_{n}$, where $x_i,y_i$ are input from $n$ parties, it is impossible to privately compute a $n$-party function $f$ if $f(\bar{x},x_{n})=f(\bar{x},y_n)=f(\bar{y},x_n)=a$ but $f(\bar{y},y_n)\neq a$ for some input. 
\end{lemma}

The above lemma is based on the private computability for two-party function cases in Kushilevitz \cite{kushelvitz1992privacy} and Chor and Kushilevitz \cite{chor1989zero}. We omit the details due to space limit.

\begin{theorem}
Without the intractability assumption, the sorting in the winner determination is not privately computable.
\end{theorem}

\begin{proof}
Lets say $f'(\bar{x},x_n)$ returns $x_n$'s rank among $x_1,\cdots,x_n$ (i.e., ranking function) and $l'(\{i_1,\cdots,i_{n-1}\},i_n)$ returns $\{i_1,\cdots,i_{n-1}\}\cup\{i_n\}$'s feasibility (i.e., feasibility check function). 

In the setting $\bar{x}=(3,1,1),\vec{y}=(5,1,1),x_4=4$ and $y_4=6$, $f'(\bar{x},x_4)=f'(\bar{x},y_4)=f'(\bar{y},x_4)=1$ but $f'(\bar{y},y_4)=2$. According to the Lemma 1, this ranking function is not privately computable. Since ranking is a necessary condition of the sorting, sorting is not privately computable either.
\end{proof}

Therefore, we design our privacy-preservin winner determination scheme based on the widely used assumption that the discrete logarithm problem is hard.

\subsection{Random Numbers in Sorting}
We assumed an infinite number domain in our framework, but in fact, all computation is conducted in a finite cyclic integer group in a real implementation. Suppose the integer group we choose is a subset of an integer group $\mathbb{Z}_p$ (i.e., $\{1,\cdots,p-1\}$) and corresponding modulo operations are followed after all arithmetic operations, then it becomes important to find `good' random numbers so that

{\footnotesize\vspace{-6pt}
\begin{displaymath}
\forall i\in \{1,\cdots,n\}:\psi_i'=\sum_{k=1}^{n}\delta_k\psi_i^{k-1}< p 
\end{displaymath}\vspace{-8pt}}

\noindent Otherwise, the rank of each bidder is not preserved after modulo operations. On the other hand, we also need to guarantee that the random numbers are large enough to securely mask all $\psi_i$'s.

%One na\"{i}ve idea is to limit the size of each random number $\delta_k, \delta_k'$ so that $\sum\delta_k \cdot \prod \delta_k'< \frac{p}{2}$ (we assume $w(i,S)\ll p$). However, this implies a necessary condition $|\delta_k'|=O(\frac{|p|}{n})$ where $|\cdot|$ stands for the big-length of the number, and $n$ is the number of total agents. This means the ciphertext length is $\Omega(n|\delta_k'|)$-bit, which is impractical for a huge $n$.

Suppose the bit lengths of integer norm $\psi_i$ is $bit_n$ and the bit lengths of every $\delta_k$ is $bit_\delta$. Then, the bit length of $\psi_i'$ is determined by the greatest term $(\delta_{n-1}+\delta_n)\psi_i^{n/2}$: 

{\footnotesize\vspace{-4pt}
\begin{displaymath}
 bit_\delta + 1 + bit_n \cdot \frac{n}{2}
\end{displaymath}\vspace{-8pt}}

\noindent which should be less than or equal to $|p|-1$ ($|p|$ is $p$'s bit length) to guarantee the correctness of the sorting. Therefore, in a real implementation, one needs to find an integer group of order $p$ of bit length $|p|$ which is at least as large as

{\footnotesize\vspace{-4pt}
\begin{displaymath}
 bit_\delta + bit_n \cdot \frac{n}{2}+ 2
\end{displaymath}\vspace{-8pt}}

\noindent Typically, $n$ is small with value at most in the order of thousands.
}

\subsection{Privacy Analysis}

\begin{theorem}
An adversarial auctioneer $Auc$'s advantage $adv_{S_i}$ is negligible.
\end{theorem}

\begin{proof}
Every winner's bundle $S_i$ is given to $Auc$, therefore we have:

{\footnotesize\vspace{-6pt}
\begin{displaymath}\begin{split}
adv_{S_i}&=\text{Pr[$S_i$}|\mathcal{C},\textsf{Output}\leftarrow \mathcal{A}_{our}(1^\kappa)]-\text{Pr[$S_i$}|\textsf{Output}\leftarrow \mathcal{A}_{black}]\\&=1-1=0
\end{split}\end{displaymath}\vspace{-11pt}}

\noindent if $\bidder_i$ is a winner.

Further, it is already shown that the Paillier cryptosystem is semantically secure \cite{paillier1999public}. Therefore, all that an adversarial $Auc$ learns during the feasibility evaluation (Section \ref{section:construction}) is whether there exists a feasible bundle whose norm is $\psi^*$. This reveals nothing about losers' $S_i$, therefore any adversary's view on losers' bundles in our algorithms is same as the one in an ideal black-box algorithm. Therefore, 

{\footnotesize\vspace{-6pt}
\begin{displaymath}
adv_{S_i}=\text{Pr[$S_i$}|\mathcal{C},\textsf{Output}\leftarrow \mathcal{A}_{our}(1^\kappa)]-\text{Pr[$S_i$}|\textsf{Output}\leftarrow \mathcal{A}_{black}]< negl(\kappa)
\end{displaymath}
}
\end{proof}
where $negl(\cdot)$ is a negligible function.

\begin{theorem}
An adversarial auctioneer's advantage $adv_{b_i}$ is negligible for all losers.
\end{theorem}

\begin{proof}
In the payment determination of the winner $\bidder_i$, a loser $\bidder_j$'s norm $\psi_j$ is disclosed to the auctioneer $Auc$, but the auctioneer does not know the identity of $\bidder_j$. In a black-box algorithm, $Auc$ gets only the output of the auction, and he learns the same norm $\psi_j$ based on $\bidder_i$'s payment $p_i=\psi_j\cdot \sqrt{|S_i|}$ and $\bidder_i$'s bundle $S_i$, but he does not know the owner of the norm either, which is a perfectly identical view as the one in our algorithms. Therefore, an adversarial $Auc$'s advantage on the candidate $\bidder_j$'s bid is negligible as well.
\end{proof}

\begin{theorem}
An adversarial bidder $\bidder_k$'s advantages $adv_{b_i}$ and $adv_{S_i}$ are all equal to negligible for all $i\neq k$.
\end{theorem}

\begin{proof}
For an adversarial bidder, he does not learn side information during our auction no matter he is a winner or not. All he learns from our privacy-preserving auction is included in the valid auction output \text{Output}. Therefore, his advantages $adv_{b_i}=adv_{S_i}< negl(\kappa)$ for all $i$, where $negl(\cdot)$ is a negligible function.
\end{proof}

\subsection{Limitation of Our Construction}

Our construction allows a winner to calculate the true payment based on the signatures. This further allows the winner to know whether the auctioneer charged more than what he deserves. But, if the winner's candidate reported a fake lower norm to the auctioneer in the first place, the winner would notice the true payment is higher than what is charged from the auctioneer, and naturally he is very likely to keep silent about this, and the auctioneer does not receive what he deserves in this case. Nevertheless, we claim this is not likely to happen due to the following reason. The candidate lost the auction because of the winner, therefore he will unlikely report a fake lower norm which is a favor of his rival.

%% file: performance_v1.tex
\section{Performance Evaluation}
\label{section:evaluation2}

\subsection{Communication Overhead}
The communication overhead in terms of the data transmission is depicted in the following table, where $n$ is the total number of bidders, $\kappa$ is the security parameter being proportional to the  order of the integer group $\mathbb{Z}_p$ and $\mathbb{Z}_n$ used in our scheme, $|W|$ is the number of winners, and $|\psi|$ is the size (bit-length) of the norm values' fixed-point representation.

\begin{table}[h]
\centering\caption{Communication Complexity}
\vspace{-3pt}
\begin{tabular}{r|c|c}
\hline\hline
\multicolumn{3}{c}{Winner Determination}\\\hline
& Receive & Send \\\hline
Auctioneer & $O(n\cdot |W| \cdot m \cdot \kappa \cdot |\psi|)$ & $O(n\cdot |W| \cdot m \cdot \kappa\cdot |\psi|)$ \\\hline
Signer & $O(n\cdot \kappa)$ & $O(n\cdot \kappa)$ \\\hline
Per bidder & $O(|W|\cdot m \cdot \kappa\cdot |\psi|)$ & $O(|W|\cdot m \cdot \kappa\cdot |\psi|)$ \\\hline\hline
\multicolumn{3}{c}{Payment Determination}\\\hline
& Receive & Send \\\hline
Auctioneer & $O(n\cdot |W|\cdot m \cdot \kappa\cdot |\psi|)$ & $O(n\cdot |W|\cdot m \cdot \kappa\cdot |\psi|)$ \\\hline
Per Winner & $O(\kappa)$ & 0 \\\hline
Per Loser & $O(|W|\cdot m \cdot \kappa\cdot |\psi|)$ & $O(|W|\cdot m \cdot \kappa\cdot |\psi|)$ \\\hline\hline
\end{tabular}\vspace{-8pt}
\end{table}

\subsection{Computation Overhead}

To evaluate the computation overhead, we implemented our privacy-preserving combinatorial auction (Section \ref{section:construction}) in Amazon EC2 using the GMP library (gmplib.org/) based on C in the c3.large instance. To exclude the communication I/O from the measurement, we generated all the communication strings (ciphertexts) and conducted all the computation in the local instance. Security parameter $\kappa$ is chosen such that the scheme enjoys 128-bit security, and every operation or protocol is run 1000 times to measured the average run time.

In general, the auction is composed of the signature generation, winner determination, and payment determination part. At each measurement (out of 1,000 times), we randomly generated the bids and bundles for all bidders and analyzed the overall run time one by one. 

\noindent \textbf{Blind Signature Generation}

The signer's run time for blindly generating one pair of the Nyberg-Rueppel signature is 5$\mu$s (microseconds) and the bidder's one is 7ms on average for each value, which are both negligible when compared to the next two.

\noindent \textbf{Winner Determination \& Payment Determination}

\begin{figure*}
  \begin{center}
    \subfigure[On different max value of $\psi$, Bidders=50, Goods=20]{\label {fig:result1-a}\includegraphics[scale=0.6]{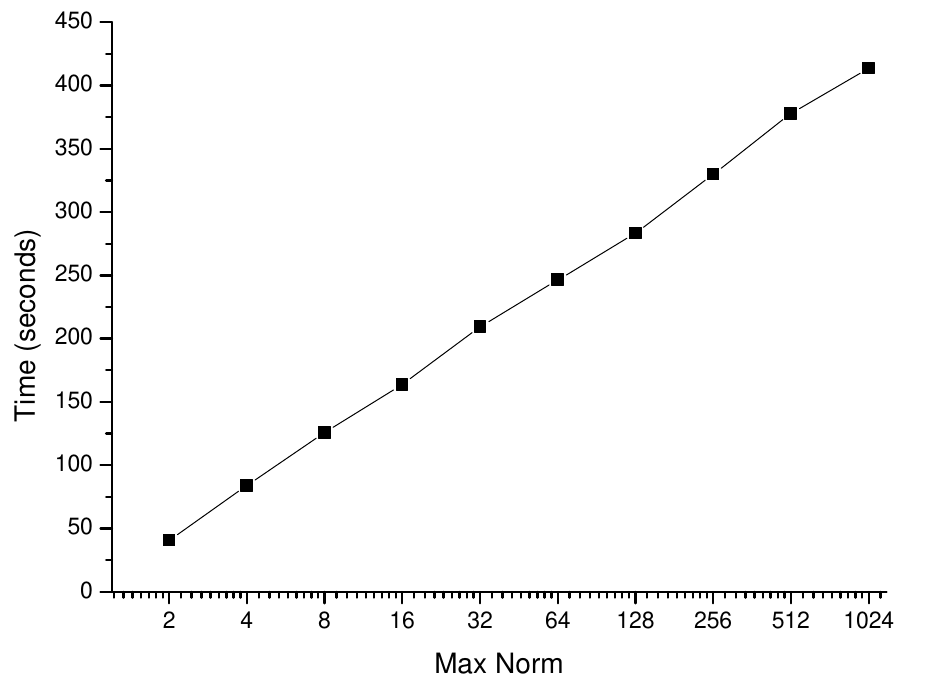}}
    \subfigure[On different total bidders, Max$\psi$=64, Goods=20]{\label {fig:result1-b}\includegraphics[scale=0.6]{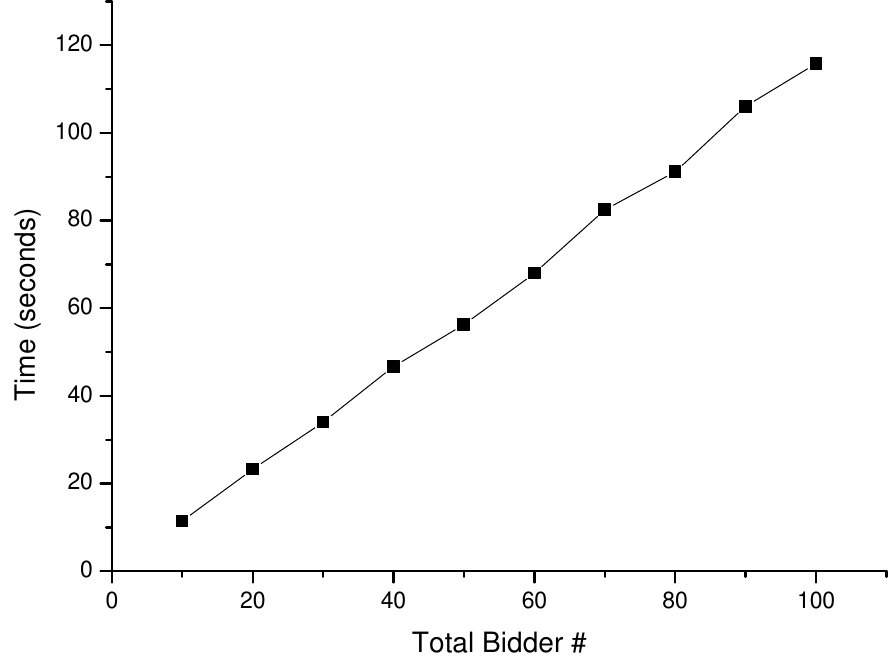}}
    \subfigure[On different total goods, Max$\psi$=64, Bidders=50]{\label {fig:result1-c}\includegraphics[scale=0.6]{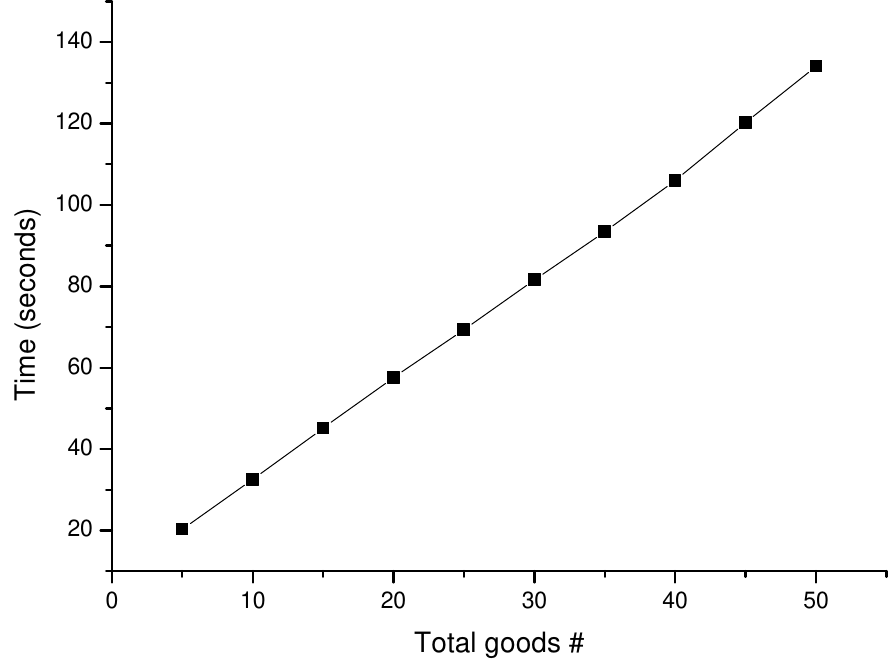}}
  \end{center}
  \vspace{-5pt}
  \caption{Auctioneer's overhead in winner determination}
  \vspace{-8pt}
  \label{fig:result1}
\end{figure*}

\begin{figure*}
  \begin{center}
    \subfigure[On different max value of $\psi$, Bidders=50, Goods=20]{\label {fig:result2-a}\includegraphics[scale=0.6]{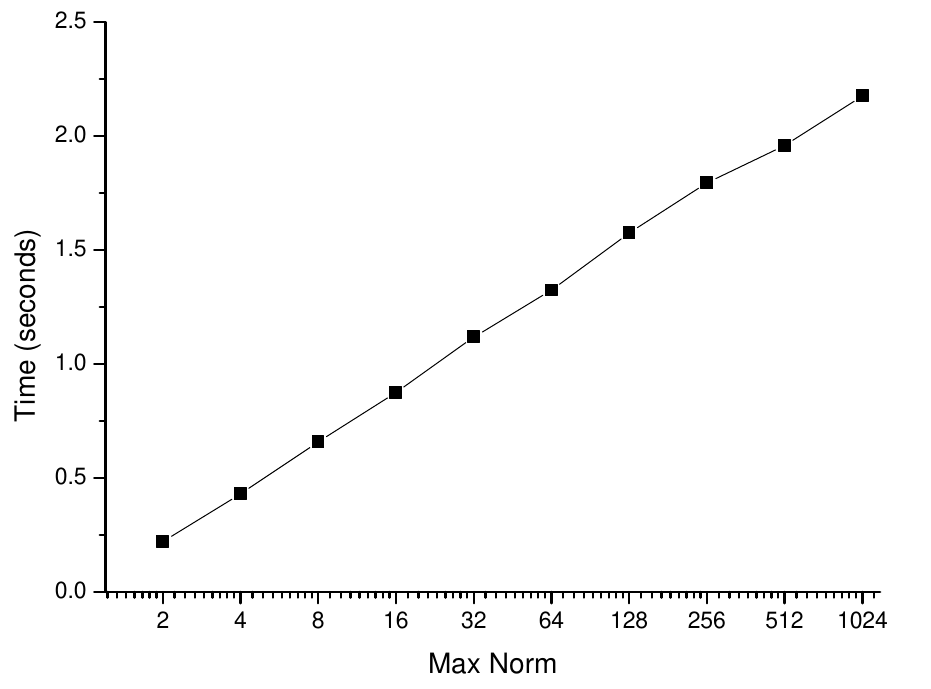}}
    \subfigure[On different total bidders, Max$\psi$=64, Goods=20]{\label {fig:result2-b}\includegraphics[scale=0.6]{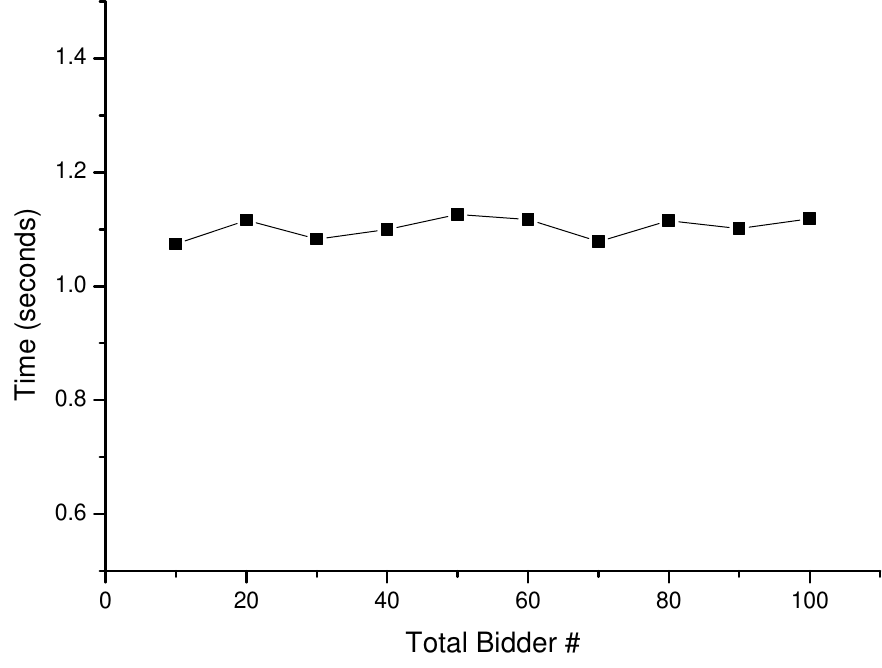}}
    \subfigure[On different total goods, Max$\psi$=64, Bidders=50]{\label {fig:result2-c}\includegraphics[scale=0.6]{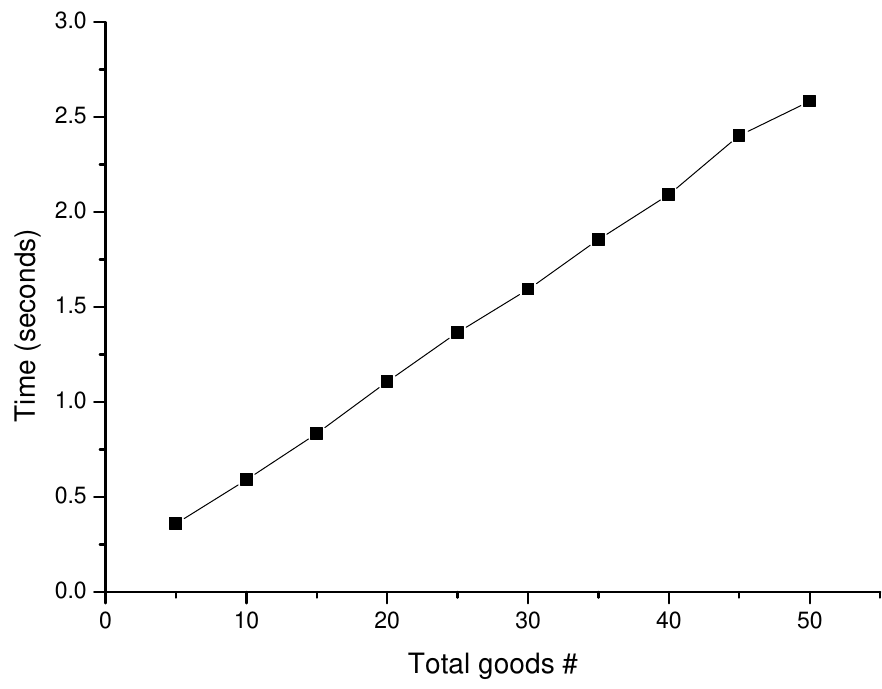}}
  \end{center}
  \vspace{-5pt}
  \caption{A bidder's overhead in winner determination}
  \vspace{-8pt}
  \label{fig:result2}
\end{figure*}

\begin{figure*}
  \begin{center}
    \subfigure[On different max value of $\psi$, Bidders=50, Goods=20]{\label {fig:result3-a}\includegraphics[scale=0.6]{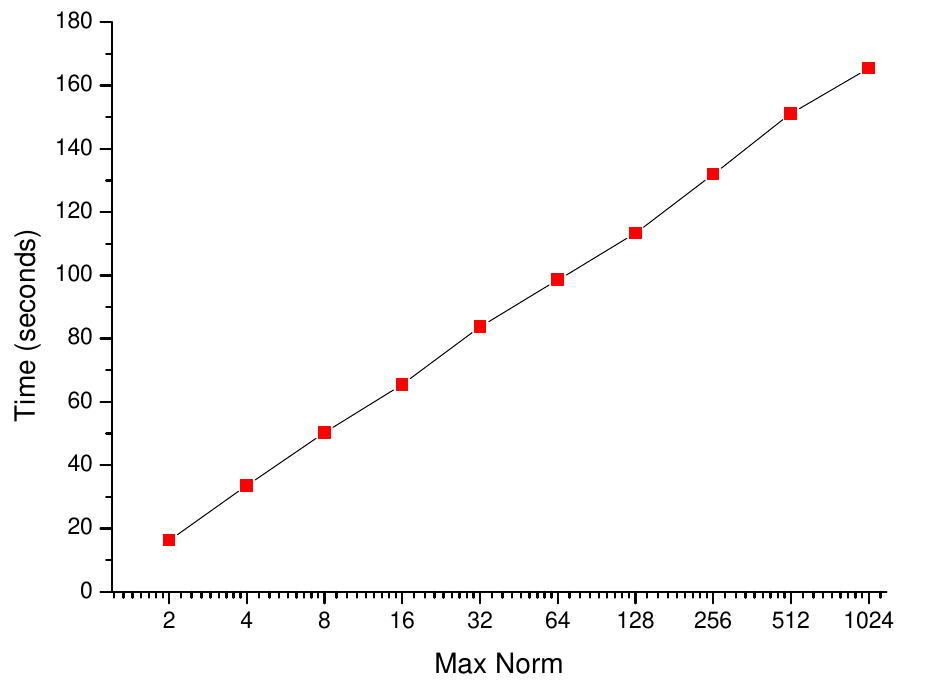}}
    \subfigure[On different total bidders, Max$\psi$=64, Goods=20]{\label {fig:result3-b}\includegraphics[scale=0.6]{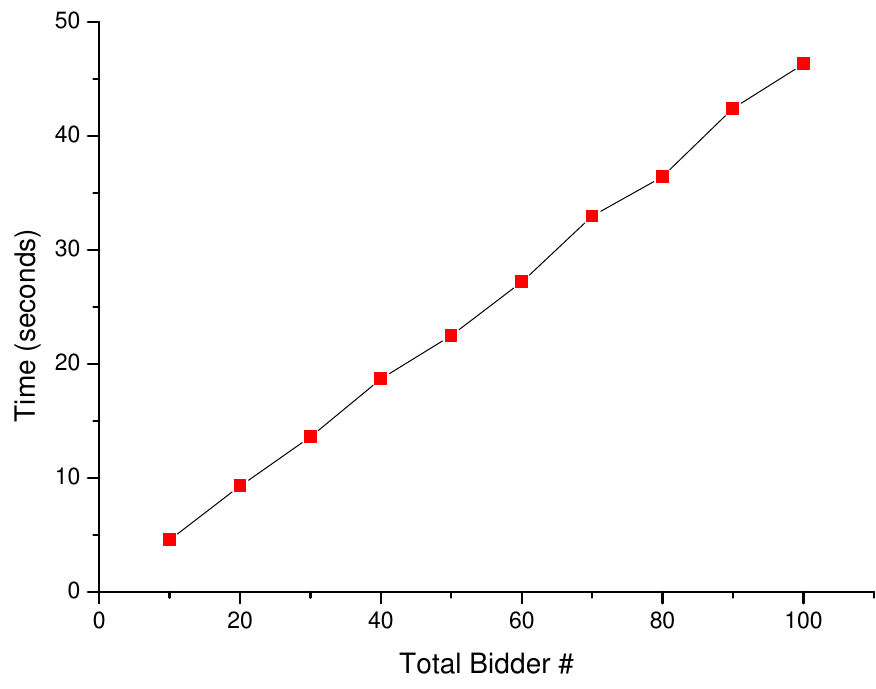}}
    \subfigure[On different total goods, Max$\psi$=64, Bidders=50]{\label {fig:result3-c}\includegraphics[scale=0.6]{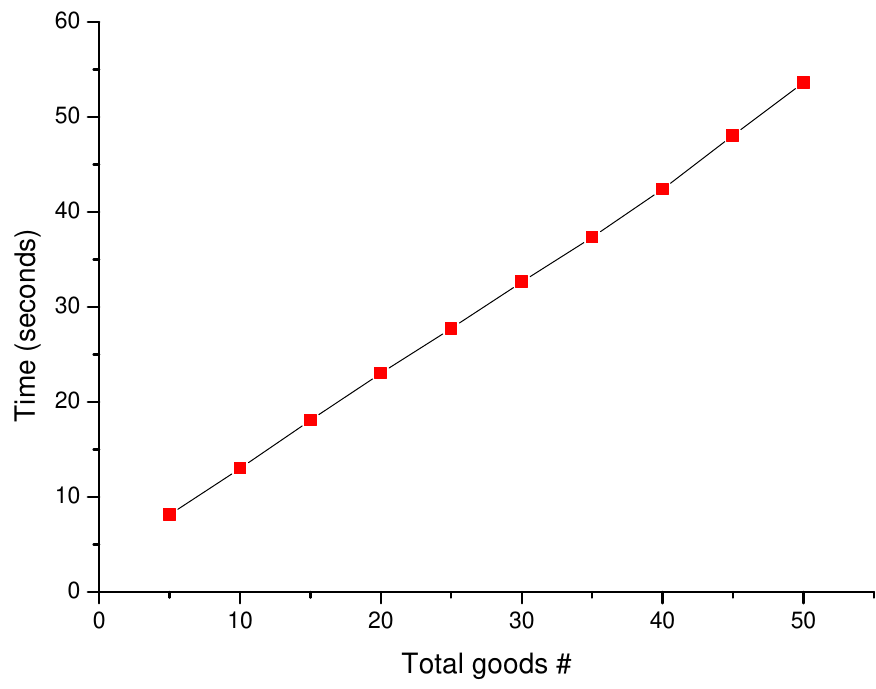}}
  \end{center}
  \vspace{-5pt}
  \caption{Auctioneer's overhead in payment determination}
  \vspace{-8pt}
  \label{fig:result3}
\end{figure*}

\begin{figure*}
  \begin{center}
    \subfigure[On different max value of $\psi$, Bidders=50, Goods=20]{\label {fig:result4-a}\includegraphics[scale=0.6]{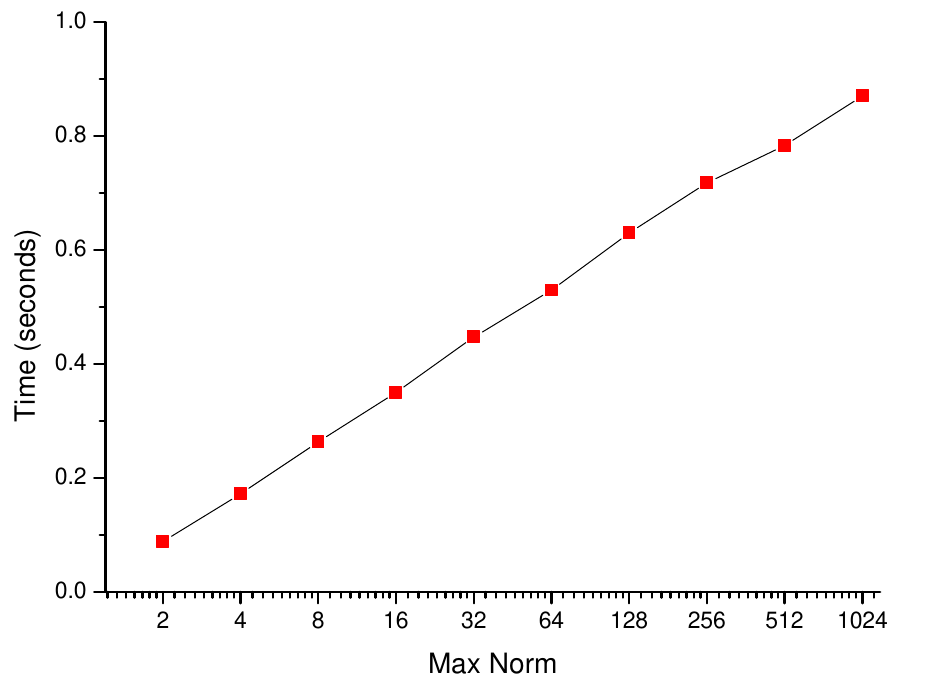}}
    \subfigure[On different total bidders, Max$\psi$=64, Goods=20]{\label {fig:result4-b}\includegraphics[scale=0.6]{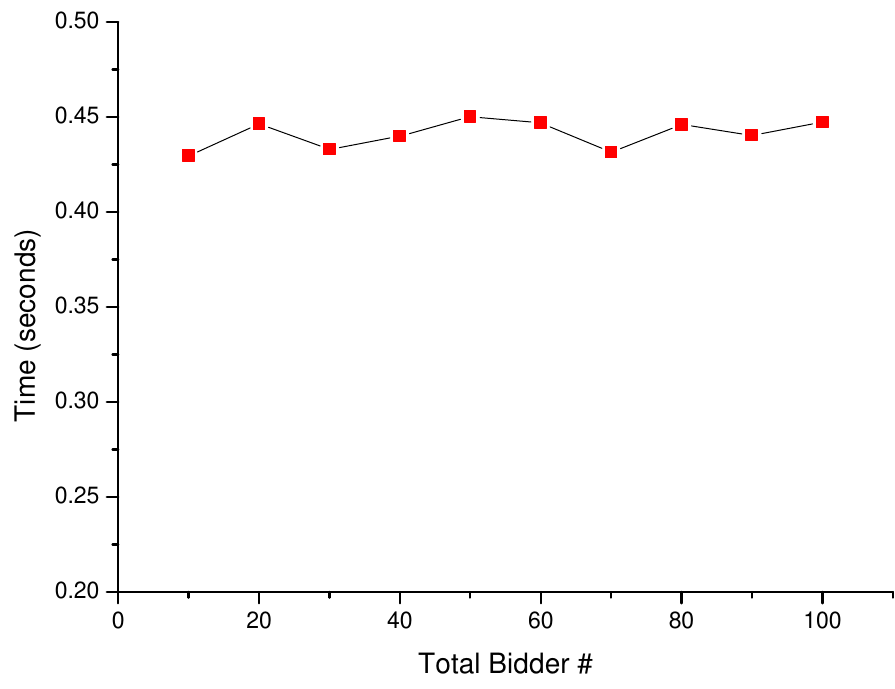}}
    \subfigure[On different total goods, Max$\psi$=64, Bidders=50]{\label {fig:result4-c}\includegraphics[scale=0.6]{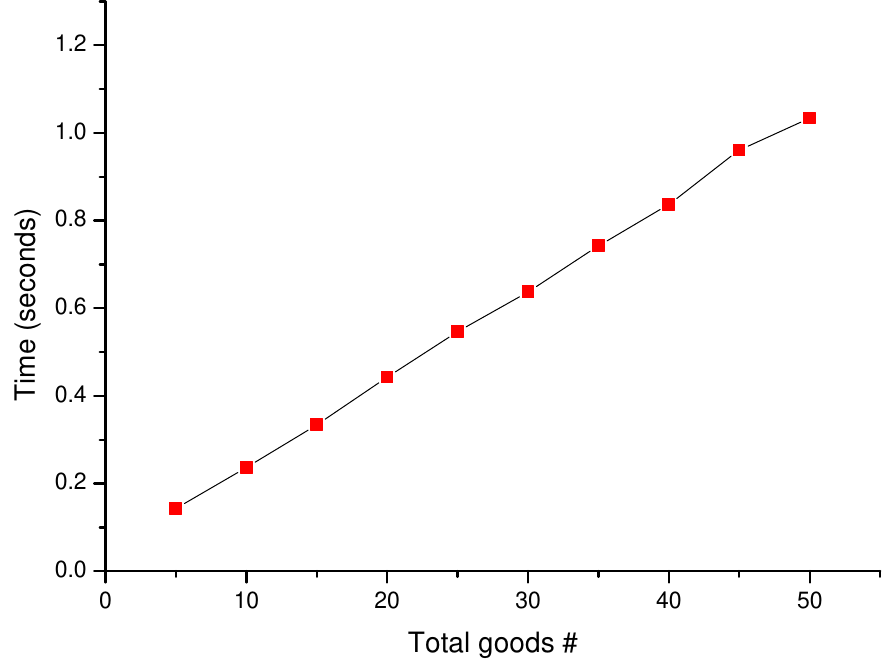}}
  \end{center}
  \vspace{-5pt}
  \caption{A bidder's overhead in payment determination}
  \vspace{-8pt}
  \label{fig:result4}
\end{figure*}

Computation overhead of winner determination and payment determination is dominated by the series of homomorphic operations, which is shown in Figure \ref{fig:result1} - Figure \ref{fig:result4}.

In fact, the homomorphic encryption we employed in this paper is a very old work proposed in 1999. Recent advances in the homomorphic encryption (\textit{e.g.,} \cite{cheon2014new}) allows much smaller overhead for the homomorphic operation, and this will improve the performance at least by 1,000 times.

\subsection{Comparison with peer works}

Besides the differences reviewed in Section \ref{section:background}, we compare scalability of ours with peer works in Table \ref{tab:scalability}.

\begin{table}[h]\label{tab:scalability}
\centering\caption{Growth of computation overhead}
\begin{tabular}{c|c|c|c}
\hline
Variable & Ours & \cite{palmer2011development} &  \cite{pan2012using,yokoo2002secure,suzuki2003secure}\\\hline
Maximum Bid & \textbf{Logarithmic} & Logarithmic & Linear\\
Bidder \# & Linear & Linear & Linear\\
Goods \# & \textbf{Linear} & Exponential & Exponential\\\hline
\end{tabular}
\end{table}

Further, by comparing the actual run time of our implementation and the one in \cite{palmer2011development}, one can notice that our overhead grows with much smaller constant factors as well. Considering that our overhead can be dramatically reduced by replacing the old Paillier's cryptosystem with more advanced and faster additive homomorphic encryption, the performance advantages over peer works  is very prominent.

%% file: conclusion_v1.tex
\section{Conclusion \& Future Direction}\label{section:conclusion}

In this paper, we presented a privacy-preserving auction design for the big data context where volume, velocity, variety, and veracity may be challenging for the auction designers. We focused on the combinatorial auction for the variety; we achieved a much better asymptotic performance than peer works by approximating the NP-hard problem in the combinatorial auction for the volume and velocity; and for the veracity issue resulted from untrusted auctioneer and bidders, we designed an auction scheme that can guarantee the truthfulness bidding and price verifiability. We extensively discussed and analyzed the security of our construction to show that any adversary's view is same as the one in a black-box algorithm, and our implementation also shows that it greatly improved the asymptotic performance of peer works. Considering the exascale computing in the big data, our work is not perfectly suitable for any big data context. However, we firmly believe this work is a big step towards the auction design in the big data era.